# Two-dimensional lattice solitons in polariton condensates with spin-orbit coupling

YAROSLAV V. KARTASHOV[1,2,*] AND DMITRY V. SKRYABIN[3,4]

[1]ICFO-Institut de Ciencies Fotoniques, The Barcelona Institute of Science and Technology, 08860 Castelldefels (Barcelona), Spain
[2]Institute of Spectroscopy, Russian Academy of Sciences, Troitsk, Moscow Region, 142190, Russia
[3]Department of Physics, University of Bath, BA2 7AY, Bath, United Kingdom
[4]ITMO University, Kronverksky Avenue 49, St. Petersburg 197101, Russia
*Corresponding author: Yaroslav.Kartashov@icfo.eu

**We study two-dimensional fundamental and vortex solitons in polariton condensates with spin-orbit coupling and Zeeman splitting evolving in square arrays of microcavity pillars. Due to repulsive excitonic nonlinearity such states are encountered in finite gaps in the spectrum of the periodic array. Spin-orbit coupling between two polarization components stemming from TE-TM energy splitting of the cavity photons acting together with Zeeman splitting lifts the degeneracy between vortex solitons with opposite topological charges and makes their density profiles different for a fixed energy. This results in formation of four distinct families of vortex solitons with topological charges $m=\pm 1$, all of which can be stable. At the same time, only two stable families of fundamental gap solitons characterized by domination of different polarization components are encountered.**

Formation of stable nonlinear excitations in the external periodic potentials is a problem attracting considerable attention in diverse areas of physics, including photonics and matter wave optics. Periodic potential may qualitatively change spatial dispersion, so that excitation of bright lattice solitons with unconventional shapes becomes possible under conditions where bright solitons simply do not exist without the potential. The properties of lattice solitons are well-studied in nonlinear optical settings, where periodic potentials are created by transverse refractive index modulations [1,2], and in Bose-Einstein condensate, where such potentials are usually formed by standing optical lattices [3].

Exciton-polaritons in microcavities represent a rapidly developing research area allowing investigation of novel aspects of light-matter interactions [4]. Among main advantages of such systems are well established technologies of the microcavity structuring allowing creation of nearly arbitrary potentials [5-9] and very strong nonlinear interactions of polaritons via their excitonic component. Nonlinear phenomena observed in the exciton-polariton condensates include superfluid behavior upon interaction with cavity defects [4], formation of oblique dark solitons and vortices [10-13], excitation of bright spatial and temporal solitons [14-21], etc. Periodic potentials created in microcavities support gap solitons in both one- [17,18] and two-dimensional [19,20] settings.

One of the most distinctive features of polariton condensate is the possibility to realize in it sufficiently strong spin-orbit coupling originating in the cavity-induced TE-TM splitting of the polariton energy levels, see, e.g. [6,8,13] and references therein. Among other effects, such a coupling mediates formation of unidirectional polaritonic edge states in truncated honeycomb potentials [8,21]. Nevertheless, the impact of spin-orbit coupling on formation of gap polariton solitons in periodic potentials was not considered yet. At the same time, it is known that in Bose-Einstein condensates spin-orbit coupling (of completely different physical origin [22]) may drastically affect stability and symmetries of available nonlinear excitations in periodic potentials [23,24].

The aim of this Letter is twofold. First, we introduce vortex polariton solitons in periodic potentials. Second, we study the impact of spin-orbit interactions, specific for polariton condensates, on all gap soliton states, including fundamental and vortex ones. In order to do this we use continuous conservative model describing evolution of exciton-polariton condensates in a square array of microcavity pillars and accounting for spin-dependent interactions and Zeeman splitting in the external magnetic field. We found that in this system the properties of vortex solitons depend on the sign of their topological charge and that four distinct stable vortex soliton families can co-exist for the same energy.

We describe the evolution of polariton condensate in periodic potential by a system of dimensionless Gross-Pitaevskii equations for the spin-positive and spin-negative components $\psi_\pm$ of the spinor wavefunction $\mathbf{\Psi} = (\psi_+, \psi_-)^{\mathrm{T}}$ [8,21]:

$$i\frac{\partial \psi_\pm}{\partial t} = -\frac{1}{2}\left(\frac{\partial^2}{\partial x^2} + \frac{\partial^2}{\partial y^2}\right)\psi_\pm + \beta\left(\frac{\partial}{\partial x} \mp i\frac{\partial}{\partial y}\right)^2 \psi_\mp + \mathcal{R}(x,y)\psi_\pm \pm \Omega\psi_\pm + (|\psi_\pm|^2 + \sigma|\psi_\mp|^2)\psi_\pm. \quad (1)$$

Here $\psi_\pm = (\psi_x \mp i\psi_y)/2^{1/2}$ are the spin-positive and spin-negative wavefunction components in the circular polarization basis that are expressed via wavefunctions $\psi_{x,y}$ associated with TM/TE cavity modes, respectively; the strength of spin-orbit coupling is described by the parameter $\beta$, which is proportional to the difference of the effective masses of the TE and TM polaritons, polarized parallel or orthogonal to the polariton momentum, see

[25]; the parameter $\Omega$ accounts for Zeeman splitting of two spin components in the external magnetic field. Thus, $\psi_\pm$ describe amplitudes of two cavity modes, bounded in $z$, but allowed to diffract in $x,y$ and evolve in time $t$. The nonlinearity stemming from exciton-exciton interactions is repulsive, but the opposite spins weakly attract, so that $\sigma=-0.05$ [26]. Due to its smallness $\sigma$ weakly affects soliton shapes. Changing its sign leads to slight soliton broadening. Microcavity pillars create periodic potential $\mathcal{R}(x,y)=-p\sum_{kl}\exp\{-[(x-ka)^2+(y-la)^2]/d^2\}$ with depth $p$, period $a$, and pillar width $d$. Quasi-conservative nonlinear dynamics has been observed in several experiments with exciton polaritons (see, e.g., [10,16]) and used in many theoretical studies (see, e.g., [8,21,25]). Following this trend, we have chosen to work here in the idealized conservative limit since the existence of soliton states is not connected with the presence of losses, while the latter will naturally limit the soliton lifetime to hundreds of ps.

All transverse coordinates in (1) are scaled to the characteristic length $x_0=1\,\mu\text{m}$, energy parameters are scaled to $\varepsilon_0=\hbar^2/2mx_0^2$, and evolution time is scaled to $t_0=\hbar\varepsilon_0^{-1}$. For the effective polariton mass $m=10^{-31}\,\text{g}$ the characteristic energy $\varepsilon_0\approx 0.35\,\text{meV}$. In this case the depth of potential $p=8$ corresponds to $2.8\,\text{meV}$, its period $a=2.2$ and pillar width $d=0.5$ correspond to $2.2\,\mu\text{m}$ and $0.5\,\mu\text{m}$, respectively, while time scale is given by $t_0\approx 1.9\,\text{ps}$.

First we consider linear spectrum of the system (1) by neglecting all nonlinear terms. The eigenmodes of this system are Bloch waves $\psi_\pm(x,y)=w_\pm(x,y)\exp(-i\varepsilon t+ik_x x+ik_y y)$, where $\varepsilon$ is the energy, $k_{x,y}$ are Bloch momenta, and complex functions $w_\pm(x,y)$ are periodic with period $a$ along both $x$ and $y$ axes. Substitution of the wavefunctions in such form into Eq. (1) yields the linear eigenvalue problem:

$$\varepsilon w_\pm = -(1/2)[(\partial/\partial x+ik_x)^2+(\partial/\partial y+ik_y)^2]w_\pm + \beta(\partial/\partial x+ik_x\mp i\partial/\partial y\pm k_y)^2 w_\mp \pm \Omega w_\pm + \mathcal{R}w_\pm, \quad (2)$$

that was solved numerically. The eigenvalues of the system (2) are arranged into bands, where periodic Bloch modes can exist. In Fig. 1 we show projections of two-dimensional dependencies $\varepsilon(k_x,k_y)$ on the $k_x$ axis. The dependencies $\varepsilon(k_x,k_y)$ are periodic in $k_{x,y}$ with the period $K=2\pi/a$. When $\beta,\Omega=0$ two equations in system (2) are decoupled. Each eigenvalue is double-degenerate, so that pairs of bands coincide. At $\beta=0$ nonzero Zeeman splitting results in the mutual shift of two initially coinciding groups of bands by $2\Omega$, as shown in Fig. 1(a). Except for this shift, the profiles of the bands do not change. It is the addition of spin-orbit coupling that leads to substantial deformation of the bands [Fig. 1(b)]. One can see that some bands (second and third from the bottom) experience notable flattening around $k_x=K/2$ point, while other bands instead broaden around it. The bands that move downwards along the $\varepsilon$ axis with increase of $\Omega$ (first and third bands from the bottom) are characterized by dominating $\psi_-$ component, while bands that move upwards (second and fourth from the bottom) correspond to wavefunctions with dominating $\psi_+$ component. Notice that the amplitude of the component that is weak in selected band gradually increases with $\beta$. In Figs. 1(a) and 1(b) one can identify sufficiently broad forbidden gap in the spectrum indicated by the line with arrows. This gap shrinks with increase of $\Omega$ (mainly due to mutual shift of the bands), so here we limit ourselves to sufficiently small value of Zeeman splitting $\Omega=0.2$. There is a general tendency that the gap and bands shrink with $\beta$.

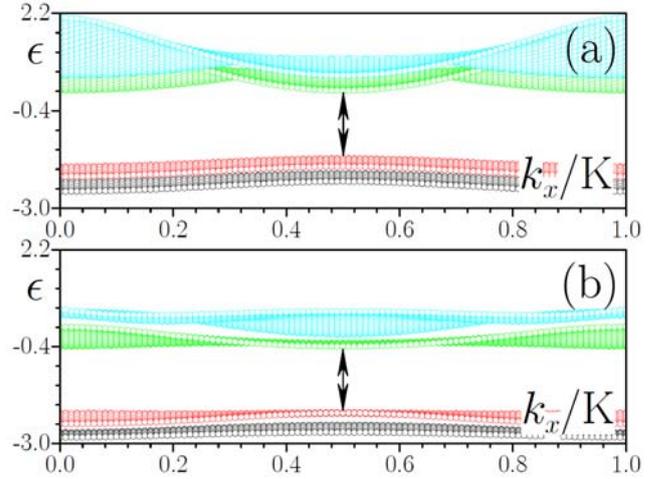

Fig. 1. (Color online) Projections of the lowest four bands in the spectrum of the square array on the $(k_x,\varepsilon)$ plane at $\beta=0$ (a) and $\beta=0.25$ (b). Line with arrows indicates finite gap where solitons are considered. In all cases $\Omega=0.2$.

Since in Eq. (1) self-repulsion dominates over weak attraction between different spin components one can find solitons in the finite gap from Fig. 1. Here we focus on fundamental and vortex gap solitons. They were obtained as stationary solutions of Eq. (1): $\psi_\pm(x,y)=w_\pm(x,y)\exp(-i\varepsilon t)$, where $w_\pm=|\psi_\pm|\exp(i\varphi_\pm)$ function describes localized soliton profile, and $\varepsilon$ is the energy that should be located in the forbidden gap. The fact of existence of two types of modes where either $\psi_+$ or $\psi_-$ wavefunction components dominate in the *linear system* finds its manifestation also in the properties of *nonlinear states*: for any value of energy $\varepsilon$ there exist two different fundamental gap solitons where either $\psi_+$ or $\psi_-$ component is much stronger. Examples of such solitons are shown in Figs. 2(a) and 3(a). We call these solitons fundamental since their strongest component [either $\psi_+$ as in Fig. 2(a), or $\psi_-$ as in Fig. 3(a)] is localized on a single pillar, at least when $\varepsilon$ falls close to the center of forbidden gap. The phase distribution in the fundamental soliton is rather complex: weak components of the wavefunction always carry *vortices*. Indeed, the structure of terms describing spin-orbit coupling in Eq. (1) is such, that while the strong $\psi_+$ component has a nearly constant phase around its maximum, the weak $\psi_-$ component coupled to it acquires phase factor $\exp(+2i\varphi)$, where $\varphi$ is the azimuthal angle. In contrast, when the strong $\psi_-$ has nearly constant phase around its maximum, then the weak $\psi_+$ varies as $\exp(-2i\varphi)$. Thus, vortices nested in different weak components have opposite topological charges. This observation is valid around density maxima of strong component. If, however, strong component $\psi_+$ features density minimum due to the presence of phase singularity $\exp(i\varphi)$ (as in vortex solitons), the weak $\psi_-$ component acquires phase $\exp(-i\varphi)$, so the sign of additional phase twist appearing in $\psi_-$ component depends also on the sign of density gradient for $\psi_+$.

It should be stressed that despite somewhat similar shapes of fundamental solitons in Figs. 2(a) and 3(a) their norms $U=\int\int_{-\infty}^{+\infty}(|\psi_+|^2+|\psi_-|^2)dxdy$ differ substantially for a fixed $\varepsilon$: a soliton with the stronger $\psi_-$ component always has larger norm [Fig. 4(a)], which will be reversed on a flip of the $\Omega$ sign. Fundamental solitons feature strongest localization when their energy $\varepsilon$ is located close to the center of the gap. The width of such solitons is a nonmonotonic function of $\varepsilon$ and it notably increases when $\varepsilon$ approaches lower or upper edges of the gap, indicated by vertical dashed lines in Fig. 4. Close to the gap edges solitons either broaden substantially or acquire small-amplitude extended background (the latter happens near the upper gap edge).

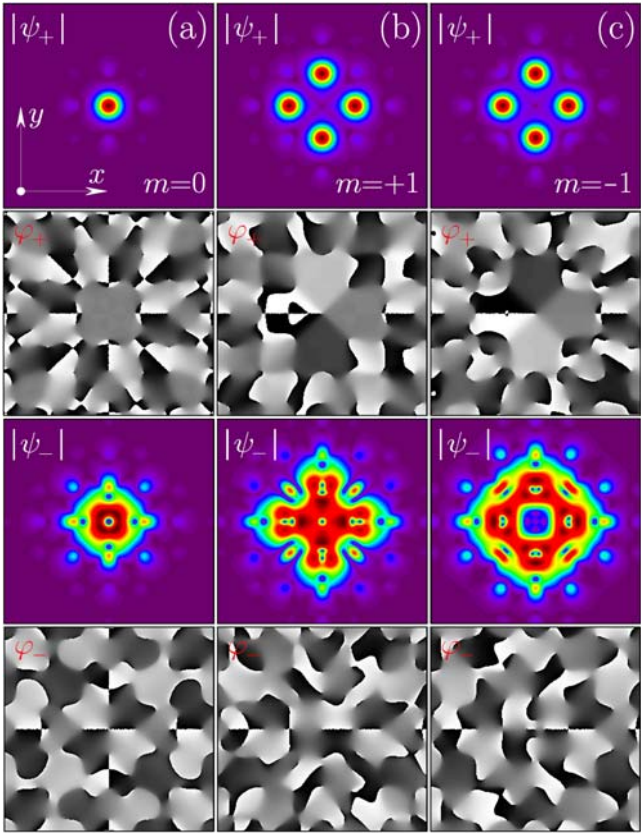

Fig. 2. (Color online) Field modulus $|\psi_\pm|$ and phase $\varphi_\pm$ distributions for fundamental gap solitons (a), as well as for on-site vortex gap solitons with topological charges $m=+1$ (b) and $m=-1$ (c) and dominating $\psi_+$ component at $\varepsilon=-1$, $\beta=0.25$, $\Omega=0.2$. All distributions are shown within $x,y\in[-7,+7]$ window.

Fundamental solitons exist above a certain threshold value of norm $U$, but this threshold is lower for the family with dominating $\psi_+$ component [Fig. 4(a)]. The latter family bifurcates exactly from the top of the red band in Fig. 1(b), hence the symmetry of solutions reflects that of the Bloch waves from this band. Peak amplitudes of components $a_\pm = \max|\psi_\pm|$ for the family with strong $\psi_+$ vanish at the lower edge of the gap [lines with solid circles in Fig. 4(b)]. Since Bloch waves in the band below gap also feature dominating $\psi_+$ component one observes that in this family $a_+ \gg a_-$ for any $\varepsilon$, even at the gap edge. In contrast, in the family with dominating $\psi_-$ component the amplitude $a_+$ grows when $\varepsilon$ approaches lower edge of the gap and becomes comparable with $a_-$ [see lines with open circles in Fig. 4(b)], again in agreement with dominance of $\psi_+$ component within the band. Even though this soliton is well localized deep in the gap, at its lower edge, where $a_+ \sim a_-$, it reshapes such that it is hard to associate it with fundamental state. Even though it can be traced up to the gap edge, we do not show this part of the family. Stability analysis performed by direct propagation of slightly perturbed solitons up to $t=10^3$ shows that only family with dominating $\psi_-$ component can be weakly unstable in small interval of energies near lower gap edge (see thick green curves). We also found that two interacting in-phase fundamental solitons placed on neighboring pillars may form stable bound states (even solitons).

Further we consider *on-site* gap vortex solitons. Examples of such states with dominating $\psi_+$ component are shown in Figs. 2(b),(c), while vortex solitons with dominating $\psi_-$ component are depicted in Figs. 3(b),(c). Largest component of the vortex soliton involves four bright spots with phase dislocation located in the center, between them, directly on the microcavity pillar.

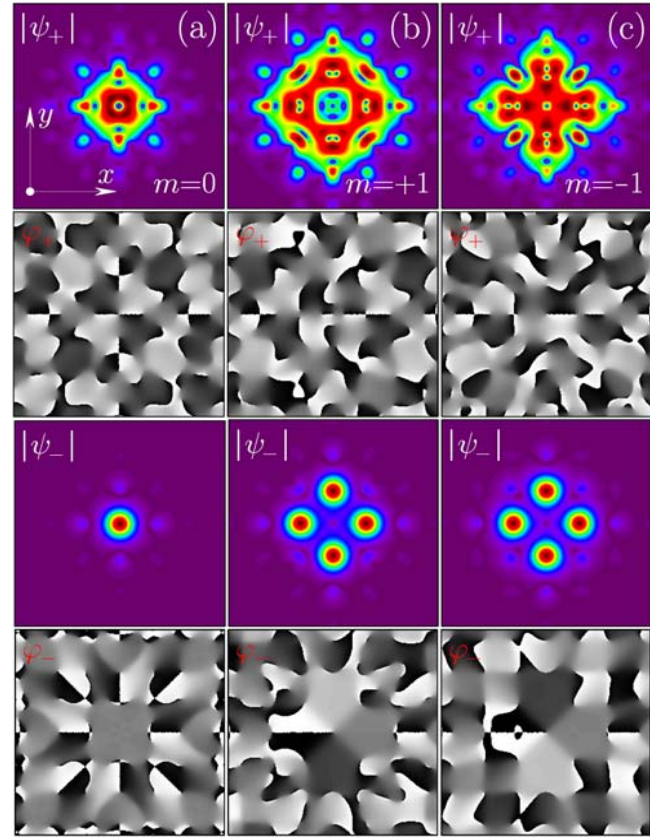

Fig. 3. (Color online) The same as in Fig. 2, but for solitons with dominating $\psi_-$ component.

Importantly, solitons with opposite topological charges $m$ in dominating $\psi_+$ components feature completely different shapes of weak $\psi_-$ components [compare Figs. 2(b) and 2(c) corresponding to $m=+1$ and $-1$, respectively]. Thus, for $m=+1$ the topological charge is inverted in the second component. We believe that for soliton with $m=-1$ [Fig. 2(c)] spin-orbit coupling tends to generate on-site phase dislocation with charge $-3$ that cannot be accommodated by the lattice due to its $C_{4v}$ discrete rotation symmetry [27]. As a result, this dislocation splits into several closely located charge-1 dislocations. All off-center charge-2 phase dislocations emerging within pillars split into two charge-1 dislocations. This splitting is different for $m=+1$ and $-1$ states.

Thus, vortex solitons with the opposite charges in the dominant $\psi_+$ field have substantially different $\psi_-$ distributions and different norms. Since a similar conclusion can be made for solutions with the dominant $\psi_-$ field [Figs. 3(b) and 3(c)], the systems under study has *four* different on-site vortex soliton families for any energy $\varepsilon$. The properties of these families are summarized in

Figs. 4(c) and 4(d) in the form of $U_\pm(\varepsilon)$ and $a_\pm(\varepsilon)$ dependencies. Note that norms of vortex solitons with opposite charges differ noticeably only close to the upper edge of the gap. Vortex solitons feature a relatively broad energy intervals, where they become weakly unstable [see branches marked with green color in Figs. 4(c) and (d)]. Solitons are unstable near the gap edges and are stable close to its center. There are energy values at which all *four* vortex families are simultaneously stable. Perturbed unstable vortices usually transform into fundamental gap solitons upon evolution.

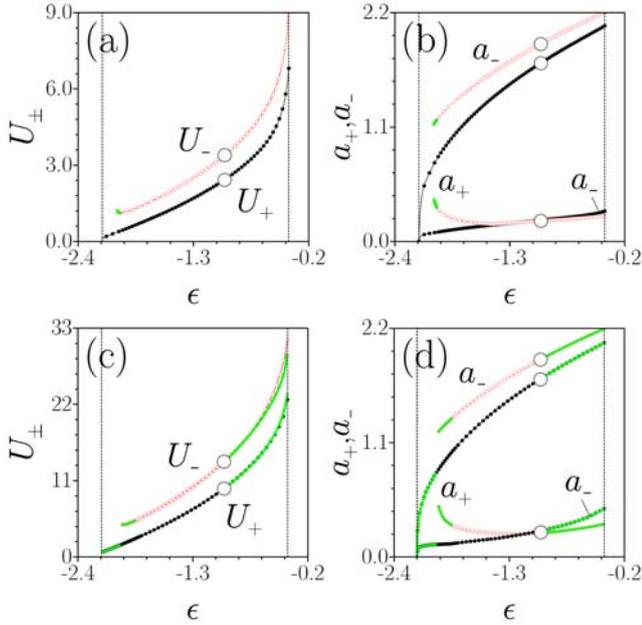

Fig. 4. (Color online) Norm (a) and peak amplitudes $a_\pm$ (b) versus energy $\varepsilon$ for fundamental solitons with dominating $\psi_+$ (lines with solid circles) or $\psi_-$ (lines with open circles) components. (c) Norm versus energy $\varepsilon$ for vortex solitons with topological charges $m=+1$ and $m=-1$. (d) Peak amplitudes $a_\pm$ of components versus $\varepsilon$ for $m=+1$ vortex state. In all plots thick green lines mark unstable branches. In (c) instability domains are indicated only for $m=+1$ states. Big white circles correspond to solitons shown in Figs. 2 and 3. Vertical dashed lines show gap edges. In all cases $\beta=0.25$, $\Omega=0.2$.

Summarizing, we have shown that spin-orbit coupling in square array of microcavity pillars lifts degeneracy between vortex solitons with opposite topological charges leading to appearance of four different vortex soliton families, all of which can be simultaneously stable. This opens a route for experimental demonstration of states whose properties depend on the sign of their topological charge.

We acknowledge support from Russian Federal Target Program (RFMEFI58715X0020), Severo Ochoa Excellence program of the Government of Spain, and Leverhulme Trust (RPG-2012-481).